\documentclass[%
 reprint,
 amsmath,amssymb,
 aps,
]{revtex4-2}
\usepackage{ulem}
\usepackage{graphicx}
\usepackage{dcolumn}
\usepackage{bm}

\usepackage{xcolor}
\usepackage{autonum}

\usepackage{comment}

\begin{document}

\preprint{APS/123-QED}

\title{No-go theorem for static configurations of two charged dust species}

\author{Andr\'es Ace\~na}
\affiliation{Instituto Interdisciplinario de Ciencias B\'asicas, CONICET, Mendoza, Argentina}
\affiliation{Facultad de Ciencias Exactas y Naturales, Universidad Nacional de Cuyo, Mendoza, Argentina}
\author{Bruno Cardin Guntsche}
\affiliation{Instituto Interdisciplinario de Ciencias B\'asicas, CONICET, Mendoza, Argentina}
\affiliation{Facultad de Ciencias Exactas y Naturales, Universidad Nacional de Cuyo, Mendoza, Argentina}
\author{Ivan Gentile de Austria}
\affiliation{Facultad de Ciencias Exactas y Naturales, Universidad Nacional de Cuyo, Mendoza, Argentina}



\begin{abstract}
We consider static spacetimes with no specific spacial symmetry where the matter content consists of two charged dust species. This comes motivated by the fact that static configurations are possible with one dust, but only if it is electrically counterpoised dust. In order to have such dust, the quotient between electric charge density and mass density needs to be fine-tuned to a value that is far less than the charge-mass quotient for any known particle. Here we prove that there are no static configurations with two dust species unless each one is electrically counterpoised dust. This shows that electrically counterpoised dust spacetimes can not be made with matter that has on average the correct charge-mass ratio, but that the underlying particles must have such ratio.
\end{abstract}

\maketitle


\section{\label{secInt}Introduction}

In the present article we consider spacetimes whose matter content is electrically counterpoised dust (ECD). Such matter corresponds to a charged perfect fluid without pressure, where the charge and mass densities are perfectly balanced. As the fluid is electrically charged, we need to consider the Einstein-Maxwell system of equations coupled to the equations of motion for the fluid itself. This may give the impression that the system of equations would turn out to be prohibitively complicated, while the opposite is true. In Newtonian Mechanics it is straightforward to see that if a collection of particles have the same mass as charge, then any static distribution is possible, as gravitational and electrostatic forces are always balanced. Strikingly, the same happens in General Relativity (GR). This was first shown for a system of discrete particles by Majumdar \cite{Majumdar1947} and Papapetrou \cite{Papapetrou1947}, following the work of Weyl \cite{Weyl1917} on axisymmetric spacetimes. If the matter content is restricted to said particles, then to each particle there is an event horizon, which is interpreted as an extremal Reissner-Nordstr\"om (ERN) black hole \cite{Hartle1972}. If instead of black holes one wants to consider regular objects, then the exterior solution can be matched with static interiors made of ECD \cite{Das1962}, \cite{Varela2003}. The reach of the results presented in \cite{Weyl1917}, \cite{Majumdar1947}, \cite{Papapetrou1947} and \cite{Das1962}, together with the minimum set of assumptions needed to obtain them, was analyzed by De and Raychaudhuri \cite{DeRaychaudhuri1968}. There, it was shown that if the spacetime is static and the matter content is dust with a constant charge to mass ratio (or if the surface of each charged dust cloud is an equipotential surface without any hole inside), then the dust is necessarily ECD and there is a particular relationship between the $tt$-component of the metric and the electrostatic potential, which by the results in \cite{Majumdar1947} imply that the metric is in fact conformastatic. The assumptions made in \cite{Weyl1917}, \cite{Majumdar1947}, \cite{Papapetrou1947}, \cite{Das1962} and \cite{DeRaychaudhuri1968} has been relaxed in several ways, and the results extended to charged perfect fluids with pressure, for example in \cite{Hernandez1967}, \cite{Guilfoyle1999}, \cite{LemosZanchin2017}, or to higher dimensions \cite{LemosZanchin2008}, \cite{LemosZanchin2009}.

Considering ECD, the fact that any static charge distribution gives rise to a solution of the Einstein-Maxwell field equations has been exploited to test features of GR, by constructing spacetimes tailored for such analysis. Therefore, properties that turned out to be difficult in a general analysis were studied in particular cases. As examples of such endeavours is the study of the relation between charge and mass in the Reissner-Nordstr\"om solution and the construction of a point charge model \cite{Bonnor1960}, the construction of static objects with unbounded density \cite{Bonnor1972}, to show that unbounded redshifts can be obtained from regular objects \cite{Bonnor1975}, and to discuss the hoop conjecture \cite{Bonnor1998}. In general, the engineered solutions can be made to be as close to the ERN black hole as desired, and this has been analyzed in relation to the bifurcation of solutions \cite{Horvat2005} and it has been shown that such black hole limit is a general feature of ECD solutions \cite{Meinel2011}. This means that a regular ECD object could mimic an ERN black hole as well as desired.

One underlying assumption when extrapolating results obtained from specific spacetimes and matter models to more general settings is that said solutions are stable. If the solution is stable one expects that physical realistic solutions close to the theoretical construction could appear in nature. If the solution is unstable then there is no expectation of finding it in nature, as it would always be subjected to some perturbations. Regarding ECD solutions, in general they are considered to possess an indifferent equilibrium, as one can go from one static distribution to another, and the system is going to remain in whichever distribution it is left. But this is true only in the sense of considering "static perturbations". For the spherically symmetric case, in the linear regime, it was shown in \cite{AcenaGentile2021} that perturbations to a static ECD solution travel at constant speed. This is a reflection of said indifferent equilibrium, but also was shown that this permits the passage from a regular solution to a black hole solution via the perturbation. Related to the question of stability of ECD is the stability of chaged fluid spheres with pressure. This problem was considered in \cite{Anninos2001}, where it was shown that in general there is a stability limit, before which the spheres are stable, and beyond it the spheres are unstable and therefore undergo gravitational collapse. In all cases the stability transition occurs before they reach the ERN limit. The combination of reaching the ERN limit and at the same time the pressure going to zero seems to be the reason why ECD ends up with an indifferent stability for static perturbations.

Another point that calls into question the feasibility of ECD solutions as physical objects, and the main motivation for the present work, is the particular fine tuning necessary between charge density and mass density. Such fine tunning is difficult to justify from more fundamental matter models. If we use geometrized units, where $G=c=1$, and $\epsilon_0 = (4\pi)^{-1}$, then the ECD condition is simply
\begin{equation}\label{sigmarho}
    \sigma=\pm\rho,
\end{equation}
where $\sigma$ is the electric charge density and $\rho$ is the mass density. If for comparison we take a gas made of protons, we have
\begin{equation}
    \frac{e}{m_p}\approx 1.1 \times 10^{18}.
\end{equation}
Therefore, if we want to construct an object of ECD with ionized hydrogen, we need to ionize a mere one in $10^{18}$ atoms. This comes from the fact that all known particles fall into two classes, in the first the particles have no electric charge and therefore the gravitational attraction can not be balanced by electric repulsion, in the second the electric repulsion is huge in comparison to the gravitational attraction. This means that there is no naturally occurring fluid where \eqref{sigmarho} is satisfied. If we want to continue with this construction, where we take a neutral gas and ionize the right proportion of atoms, then we are forced to consider the presence of two species. From the previous example, one is neutral hydrogen, with no charge to balance the gravitational pull, and the other is ionized hydrogen, with charge density much higher than required. This argument leads to the consideration of two charged dust species, to see if it is possible within GR to construct configurations where the required relationship \eqref{sigmarho} is satisfied only on average and not for each fluid species separately. If this is not possible, as we prove here, then we consider that there is no natural situation where ECD could be expected to occur.

The article is organized as follows. In Section \ref{SecTheo} we state the problem and present the result. The proof of the no-go theorem is developed in Section \ref{SecProof}, followed by the conclusions in Section \ref{SecConc}.

\section{\label{SecTheo}Problem statement and no-go theorem}

We consider a static spacetime where the matter content are two electrically charged dust species, which we denote by $A$ and $B$. The proper energy density of the first fluid is $\rho_A$, and its proper electric charge density is $\sigma_A$. Respectively for the second fluid we have $\rho_B$ and $\sigma_B$. We use coordinates adapted to the staticity of the spacetime, $(t,x,y,z)$, but assume no spacial symmetry. Therefore, the named densities are functions of $(x,y,z)$. Following \cite{DeRaychaudhuri1968}, we further assume that the ratio of total matter density to total charge density is constant, in our setting this means that
\begin{equation}\label{assumption}
    \frac{\sigma}{\rho} = \frac{\sigma_A+\sigma_B}{\rho_A+\rho_B} = constant.
\end{equation}
This assumption, while technical, is justified by the fact that what we are trying to test is the existence of solutions that satisfy \eqref{sigmarho}, that is, test the existence of solutions that are on average ECD. Although we have this motivation, we do not need to assume in \eqref{assumption} that the ratio is that of ECD. The theorem is stated as:

\vspace{5pt}
\textit{\textbf{No-go theorem: }There is no static spacetime with two electrically charged dust species that satisfy \eqref{assumption} unless
\begin{equation}\label{eqTheo}
    \sigma_A = \pm\rho_A\quad\mbox{and}\quad \sigma_B = \pm\rho_B.
\end{equation}}

Please note that the same sign needs to be chosen in the equalities \eqref{eqTheo}, as both species have to repel each other to balance the gravitational attraction. Also, the assumption \eqref{assumption} can be replaced by the assumption that the surface of each dust cloud is an equipotential surface without any hole inside, but we prefer \eqref{assumption} as our intention is to test the feasibility of ECD solutions. If \eqref{eqTheo} are satisfied, then the two species can not be distinguished by their mass-charge ratio, and therefore they are effectively only one fluid for the setting at hand.

\section{\label{SecProof}Proof of the theorem}

We need to consider the Einstein-Maxwell system of equations together with the equations of motion for each dust species. The Einstein equations are
\begin{equation}
    G_{\mu\nu} := R_{\mu\nu}-\frac{1}{2}g_{\mu\nu}R = 8\pi T_{\mu\nu},
\end{equation}
where the energy-momentum tensor has contributions from the fluids as well as from the electromagnetic field,
\begin{equation}
    T_{\mu\nu} = T_{\mu\nu}^A + T_{\mu\nu}^B + T_{\mu\nu}^{EM}.
\end{equation}
If we denote by $u_\mu^A$ and $u_\mu^B$ the four-velocities of each fluid species, then, as they are both dusts,
\begin{equation}
   T_{\mu\nu}^A = \rho_A u_\mu^A u_\nu^A,\quad T_{\mu\nu}^B = \rho_B u_\mu^B u_\nu^B.
\end{equation}
The electromagnetic contribution is
\begin{equation}
    T^{EM}_{\mu\nu} = \frac{1}{4\pi} \left(F_{\gamma\mu} F^{\gamma}\,_{\nu}-\frac{1}{4}F_{\gamma\lambda}F^{\gamma\lambda}g_{\mu\nu}\right),
\end{equation}
where the Faraday tensor, $F_{\mu\nu}$, is written in terms of the electromagnetic four-potential, $A_\mu$,
\begin{equation}
    F_{\mu\nu} = \nabla_\mu A_\nu - \nabla_\nu A_\mu.
\end{equation}
The Maxwell equations are
\begin{equation}\label{eqAgr}
    \nabla_\nu F^{\mu\nu} = 4\pi j^\mu,
\end{equation}
where $j^\mu$ is the current density. With the previous notation we have
\begin{equation}
    j^\mu = \sigma_A u_A^\mu + \sigma_B u_B^\mu.
\end{equation}
The equations of motion for the dust species are
\begin{equation}\label{eom}
  \rho_A u^\nu_A\nabla_\nu u_A^\mu = f^\mu_A,\quad\rho_B u^\nu_B\nabla_\nu u_B^\mu = f^\mu_B,
\end{equation}
where $f^\mu_A$ and $f^\mu_B$ are the Lorentz forces on each fluid,
\begin{equation}
    f_A^\mu = \sigma_A F^{\mu\nu}u_\nu^A,\quad f_B^\mu = \sigma_B F^{\mu\nu}u_\nu^B.
\end{equation}

Now we restrict to the static case. The metric can be written as
\begin{equation}
    ds^2 = g_{tt} dt^2 + g_{ij} dx^i dx^j
\end{equation}
where $g_{tt}$ and $g_{ij}$ are functions of the spatial coordinates, $x^i$, only. Due to staticity, the four-velocities are
\begin{equation}
    u^\mu_A = u^\mu_B = (-g_{tt})^{-\frac{1}{2}}\,\partial_t,
\end{equation}
and there is an electrostatic potential, $V(x^i)$, whith which the electromagnetic four-potential takes the form
\begin{equation}
    A_\mu = V\,dt.
\end{equation}
In \cite{DeRaychaudhuri1968} it is proven that given the Einstein-Maxwell equations plus the hypothesis that \eqref{assumption} is satisfied then there is a functional relationship between $g_{tt}$ and $V$, which can be written as
\begin{equation}
    g_{tt} = -V^2,
\end{equation}
and that $\sigma=\pm\rho$. Furthermore, by the results in \cite{Majumdar1947}, the metric is conformastatic, which means that it can be written in the form
\begin{equation}
    ds^2 = -H^{-2} dt^2 + H^2 (dx^2+dy^2+dz^2),
\end{equation}
where $H$ is a function of $(x,y,z)$. These results carry over to the case at hand directly with $\rho=\rho_A+\rho_B$ and $\sigma=\sigma_A+\sigma_B$. It is then concluded that the Einstein-Maxwell equations 
reduce to a single equation for $H$,
\begin{equation}\label{ecH}
    \Delta H = - 4\pi H^3 (\rho_A + \rho_B),
\end{equation}
where $\Delta$ is the flat Laplacian. Therefore, given $\rho_A$ and $\rho_B$, \eqref{ecH} is solved for $H$, the electromagnetic field is obtained through
\begin{equation}\label{ecV}
    V = \pm H^{-1}
\end{equation}
and the charge densities have to satisfy
\begin{equation}\label{ecS}
    \sigma_A + \sigma_B = \pm(\rho_A+\rho_B),
\end{equation}
where the same sign needs to be chosen in these last two equations. Here it may seem that the ECD condition can be satisfied on average. This is due to the fact that in the energy-momentum tensor all dust species contributions enter in the same way, and therefore from the energy-momentum tensor it is not possible to disentangle to which species corresponds a particular matter density contribution. The same happens regarding the charge contributions in the Maxwell equations. As a consequence we have that from the Einstein-Maxwell equations alone it is not possible to find the equations of motion for the fluids. These are \eqref{eom} and for the case at hand they give
\begin{equation}
    -\rho_A \frac{\partial_i H}{H^2} = \sigma_A \partial_iV,\quad -\rho_B \frac{\partial_i H}{H^2} = \sigma_B \partial_iV,
\end{equation}
where the index $i$ stands for $x$, $y$ and $z$. By \eqref{ecV} this simply reduces to
\begin{equation}
    \sigma_A = \pm \rho_A,\quad \sigma_B=\pm\rho_B,
\end{equation}
which means that both dust species have to be ECD. Here the sign needs to coincide with the one in \eqref{ecV}. This concludes the proof.

\section{\label{SecConc}Conclusions}

We have considered spacetimes where the matter content is given by two electrically charged dust species and shown that there are no static distributions unless each one is individually ECD. This implies that it is not possible to form static ECD objects using charged dust species that do not satisfy \eqref{sigmarho} and averaging the mass and charge densities. The generalization of the result to more than two species is straightforward. Also, given that no known particle satisfies \eqref{sigmarho}, being the charge and mass not balanced by orders of magnitude, then static distributions of charged dust are not expected to occur.

The present result adds to the still open question posed in \cite{Anninos2001}: "Can relativistic charged spheres form extremal black holes?". This question is also part of a more general discussion concerning the cosmic censorship conjecture, where extremal black holes are seen as the way of passing (or being an impassable barrier) between black holes and naked singularities. Starting a fruitful discussion in \cite{Wald1974}, it was shown that Kerr-Newman black holes can not be overspun or overcharged by test particles. This was extended to a plethora of more general settings, for example in \cite{Natario2016} and \cite{Sorce2017}. These works strongly indicate that to form an extremal black hole by gravitational collapse it is necessary to start with a distribution of matter which is already extremal. In \cite{Anninos2001} it was shown that before this extremal matter limit is attained the object undergoes gravitational collapse, which strongly suggests that ERN black holes can not be produced by the collapse of charged spheres. Although less physically reasonable from the point of view of its equation of state, ECD is the natural candidate to form an ERN black hole by collapse, being the relationship \eqref{sigmarho} the microscopic equivalent of the extremality condition $Q=M$. In \cite{AcenaGentile2021} it was shown that indeed, in the linear perturbations regime, an ERN black hole can be formed, wich indicates that if ECD spacetimes were possible, then the formation of ERN black holes would be a reality. Opposing this, the present result shows that unless there is a particle with the correct charge-mass ratio to start with then ECD spacetimes are not possible.

\begin{acknowledgments}
Part of the calculations were performed using SageMath \cite{SageMath} with the package SageManifolds \cite{SageManifolds}.

This work was partially supported by grants M001-T1 and M020-T1 of SIIP, Universidad Nacional de Cuyo, Argentina.
\end{acknowledgments}





\end{document}